\documentclass[aip, jap, reprint]{revtex4-1}
\pdfoutput=1
\usepackage{amsmath}
\usepackage[colorinlistoftodos]{todonotes}
\usepackage{graphicx}
\usepackage{float}
\usepackage{hyperref}

\newcommand{\figwit}[0]{0.95\textwidth}
\begin{document}
\title{Skyrmion states in thin confined polygonal nanostructures}
\author{Ryan Alexander Pepper}
\email{ryan.pepper@soton.ac.uk}
\affiliation{Faculty of Engineering and the Environment, University of
  Southampton, Southampton SO17 1BJ, United Kingdom}
\author{Marijan Beg}
\affiliation{Faculty of Engineering and the Environment, University of
  Southampton, Southampton SO17 1BJ, United Kingdom}
\affiliation{European XFEL GmbH, Holzkoppel 4, 22869 Schenefeld, Germany}
\author{David Cort\'{e}s-Ortu\~{n}o}
\author{Thomas Kluyver}
\author{Marc-Antonio Bisotti}
\author{Rebecca Carey}
\author{Mark Vousden}
\author{Maximilian Albert}
\affiliation{Faculty of Engineering and
the Environment, University of Southampton, Southampton SO17 1BJ,
United Kingdom}
\author{Weiwei Wang}
\affiliation{Department of Physics, Ningbo University, Ningbo 315211, China}
\author{Ondrej Hovorka}
\affiliation{Faculty of Engineering and
the Environment, University of Southampton, Southampton SO17 1BJ,
United Kingdom}
\author{Hans Fangohr}
\email{hans.fangohr@xfel.eu}
\affiliation{Faculty of Engineering and
the Environment, University of Southampton, Southampton SO17 1BJ,
United Kingdom}
\affiliation{European XFEL GmbH, Holzkoppel 4, 22869 Schenefeld, Germany}
\begin{abstract}
  Recent studies have demonstrated that skyrmionic states can be the ground
  state in thin-film FeGe disk nanostructures in the absence of a stabilising
  applied magnetic field. In this work, we advance this understanding by
  investigating to what extent this stabilisation of skyrmionic structures
  through confinement exists in geometries that do not match the cylindrical
  symmetry of the skyrmion -- such as as squares and triangles. Using
  simulation, we show that skyrmionic states can form the ground state for a
  range of system sizes in both triangular and square-shaped FeGe nanostructures of $10\,\text{nm}$
  thickness in the absence of an applied field. We further provide data to assist in the experimental
  verification of our prediction; to imitate an experiment where the system is
  saturated with a strong applied field before the field is removed, we compute
  the time evolution and show the final equilibrium configuration of
  magnetization fields, starting from a uniform alignment.
\end{abstract}
\maketitle
\section{Introduction}\label{introduction}
Magnetic skyrmions have been an active research area in recent years after
theoretical predictions of formation in materials with broken inversion
symmetry, which host a Dzyaloshinskii-Moriya (DM) interaction.
\cite{Bogdanov1994, Bogdanov1999, Roszler2006} These predictions have been
experimentally realised in a variety of materials, such as in the bulk metallic
cubic B20 materials FeGe \cite{Muhlbauer2009} and MnSi \cite{Lee2009,
Pfleiderer2009, Neubauer2009a, Jonietz2010}, the insulating
$\text{Cu}_2\text{OSeO}_3$ \cite{Zhang2016g}, and in thin film and multilayered
systems. \cite{Crepieux1998, Heinze2011, Moreau-Luchaire2016} Driving this
research, aside from the interest in the physics of such systems, are potential
engineering applications to data storage and logic devices. The application to
data storage in particular is important due to current challenges in existing
technology. The magnetic recording trilemma \cite{Richter2007} is a well known
problem with domain based storage, whereby the shrinking of current domain sizes
competes with potential data loss from thermal fluctuations, requiring
magnetically stiff materials and a correspondingly higher write field, which
becomes difficult to achieve. Magnetic skyrmions, which can be of a much smaller
size than the current domains, are a potential solution to this problem because
the topology of the magnetization can provide a larger energy barrier to
destruction. \cite{Fert2013, Bessarab2015a, Cortes-Ortuno2016} Experiments have
shown that skyrmion creation and deletion can be achieved through the injection
of spin polarised currents, and that skyrmion manipulation can be achieved with
low current densities relative to magnetic domain walls. \cite{Romming2013,
Jonietz2010, Yu2012} Storage device proposals include racetrack based storage,
where the presence or absence of a skyrmion system can represent a bit. \cite{Fert2013, Zhang2015i}

Recent studies have shown nucleation of skyrmions at room temperature in bulk
systems and in interfacial systems, which brings the goal of creating devices
much closer. \cite{Woo2016, Krause2016} The physics and energetics of confined
geometries differ significantly from those of large bulk
systems. \cite{Rohart2013} This is of particular concern for magnetic
systems because topological protection is not afforded to skyrmions in
finite-sized systems; skyrmions can be destroyed via variation of the
magnetization field at the boundary, with a significantly lower energy barrier
than other skyrmion destruction mechanisms.\cite{Cortes-Ortuno2016, Uzdin2017} To this end,
it is important to understand how the confined nature of the geometry can affect
the energetics of the skyrmion states. In a previous study, FeGe nanodisks were
studied through micromagnetic simulations, and it was found that skyrmion states
could form the ground state in a narrow range of disk sizes, with no applied
magnetic field. \cite{Beg2015} This demonstrates a stabilisation of the skyrmion
via the sample boundary. Recently, experimentalists have createad FeGe
nanodisks, and have observed skyrmion cluster and target states, \cite{Zhao2016,
Zheng2017} in line with theoretical predictions \cite{Beg2015, Carey2016b}. However, it is not obvious that these results can be extended to
other geometries, as the boundary of the system has a significant effect on the
magnetization. In this paper we advance the understanding of skyrmions in
confined geometries, by studying polygonal films of FeGe and investigate the
ground and metastable states of these systems for a range of sizes and applied
fields. We choose regular polygonal films in order to study how the shape of the
systems affects the equilibrium states which form, and how this changes the
lowest energy magnetization configuration at each system size.

\section{Method}
We study, through micromagnetic simulations, film systems of FeGe of thickness
$10\,\text{nm}$ using a fully three-dimensional model. This model is chosen as
it has been rigorously shown both theoretically and experimentally that in films
of cubic helimagnets, chiral modulations occur along all three spatial
dimensions, which reduces the skyrmion state energy in 3D systems of thickness
lower than the helical length. \cite{Rybakov2016b, Vousden2016, Leonov2016b, Schneider2017}
The dynamics of the magnetization field $\mathbf{m}$ are modelled by the
Landau-Lifshitz-Gilbert (LLG) equation
\begin{equation}
\frac{\partial \textbf{m}}{\partial t} = \gamma_0^* \textbf{m} \times \textbf{H}_\text{eff} + \alpha \textbf{m} \times \frac{\partial\textbf{m}}{\partial t}.
\end{equation}
Here, $\gamma_0^*=\gamma_0\left(1 + \alpha^2\right)$ where $\gamma_0$ is the
gyromagnetic ratio, and $\gamma_0 < 0$. The constant $\alpha$ is the Gilbert
damping coefficient. The effective magnetic field is calculated as
$\mathbf{H}_\text{eff} = - \left(\delta w / \delta \textbf m\right) / (\mu_0
M_s)$, where $w$ is the total energy density given as:
\begin{equation}
w = w_\text{Exchange} + w_\text{DM} + w_\text{Zeeman} + w_\text{Demag}
\end{equation}
The symmetric exchange energy density is $w_\text{Exchange} =
A\left(\nabla\textbf{m}\right)^2$ where $A$ is the magnetic exchange constant.
The bulk Dzyaloshinskii-Moriya interaction (or antisymmetric exchange) in a
material of crystallographic class T is given as $w_\text{DM} = D\textbf{m}
\cdot \left( \nabla \times \textbf{m} \right)$ where $D$ is the DMI energy
constant. The Zeeman energy is calculated from the applied field $\mathbf{H}$ as
$w_\text{Zeeman} = -\mu_0 M_\text{s}\mathbf{m}\cdot\mathbf{H}$. The
demagnetizing field is calculated using the Fredkin-Koehler hybrid FEM/BEM
method.\cite{Fredkin1990} For the simulations of FeGe, we use the parameters
\cite{Beg2015} $A = 8.78\times 10^{-12}\,\text{J m}^{-1}$, $D = 1.58\times
10^{-3}\,\text{J m}^{-2}$, $M_s = 3.84\times 10^{5}\text{A m}^{-1}$. The
finite-element discretisation was set such that the distance between mesh nodes was no greater than $3\,\text{nm}$, smaller than the relevant micromagnetic length scales for the given material, which has a helical length of $70\,\text{nm}$ and exchange length $l_\text{ex} = \sqrt{\frac{2A}{\mu_0 M_s^2}} = 9.67\,\text{nm}$

\begin{figure}
\centering
\includegraphics[width=\linewidth]{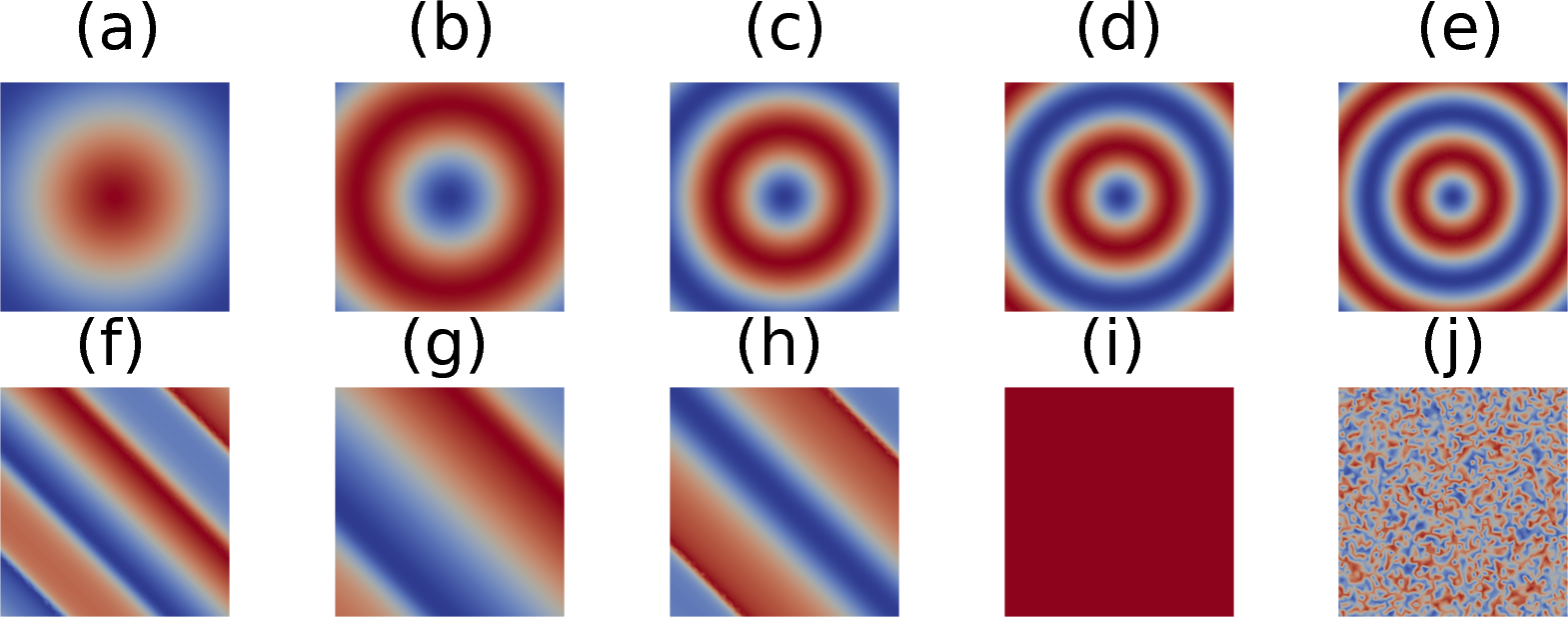}
\caption{Initial magnetization configurations from which each geometry is relaxed,
shown here in a 140$\,$nm side length film. The states are (a)~incomplete
skyrmion, (b)~isolated skyrmion, (c) and (d) overcomplete skyrmions (e) target state
(f), (g) and (h) helical states of different helical lengths (i) uniform
state, and (j)~random state.}
\label{fig:initialstates}
\end{figure}

We compute the ground state phase diagram for two types of FeGe sample; square
and triangular films of $10 \,\text{nm}$ thickness, through dynamic simulations.
Dynamic simulations are used in order that all discovered states are physically
realisable. We explore the energy landscape of structures by changing the
applied magnetic field, which is varied between $0\,\text{mT}$ and
$800\,\text{mT}$, and which is applied in the $z$ direction, into the plane of
the system. In squares, we study films which have a side length of between
$40\,\text{nm}$ and $180\,\text {nm}$, and in triangles we use the broader range
of side lengths between $40\,\text{nm}$ and $220\,\text{nm}$. Initially, the
magnetization of each point in the phase space is set to each configuration of a
set of initial states; the definition of these states is the same as those used
in the study of Beg et al. \cite{Beg2015} (see Supplementary Material). The set of initial states, which
includes uniform magnetization, skyrmionic state profiles, helical profiles,
and a random magnetization state (which is repeated three times) are shown in
Fig.~\ref{fig:initialstates}. This systematic exploration is done in order to
capture as many equilibrium states as possible for each simulated system. In
order to construct the ground state phase diagrams, we relax systems from these
initial states under the LLG equation, until the system has settled into a local
(or global) minima in the energy landscape. States are considered to be in
equilibrium, and simulations are stopped, when the value of $\lvert\partial
\mathbf{m}/\partial t\rvert$ is less than a tolerance of 0.01 degrees per
nanosecond, at which point the magnetization is no longer changing. We use a
damping factor value of $\alpha = 1$ in order to achieve convergence to the
final states quickly, by suppressing the precessional dynamics, which does not
affect the final state. Once the dynamics have subsided according to the above
criterion, we compute the total energy of these relaxed states. We identify the
lowest energy state that we have found (from the set of simulations starting
from different initial configurations) as the ground state for the given
geometry and applied field value, which allows us to construct $d - B$ phase
diagrams of the ground states. Higher energy states we consider to be
metastable.

To perform the simulations, we use the finite-element micromagnetic
simulator Finmag, developed at the University of Southampton. This uses the
DOLFIN component of the finite-element solver FEniCS,\cite{Logg2010} and
integrators from the CVODE component of the SUNDIALS library.\cite{Hindmarsh2005}
\section{Results}\label{study}
\subsection{Equilibrium States}
\begin{figure*}
\centering
\includegraphics[width=0.89\textwidth]{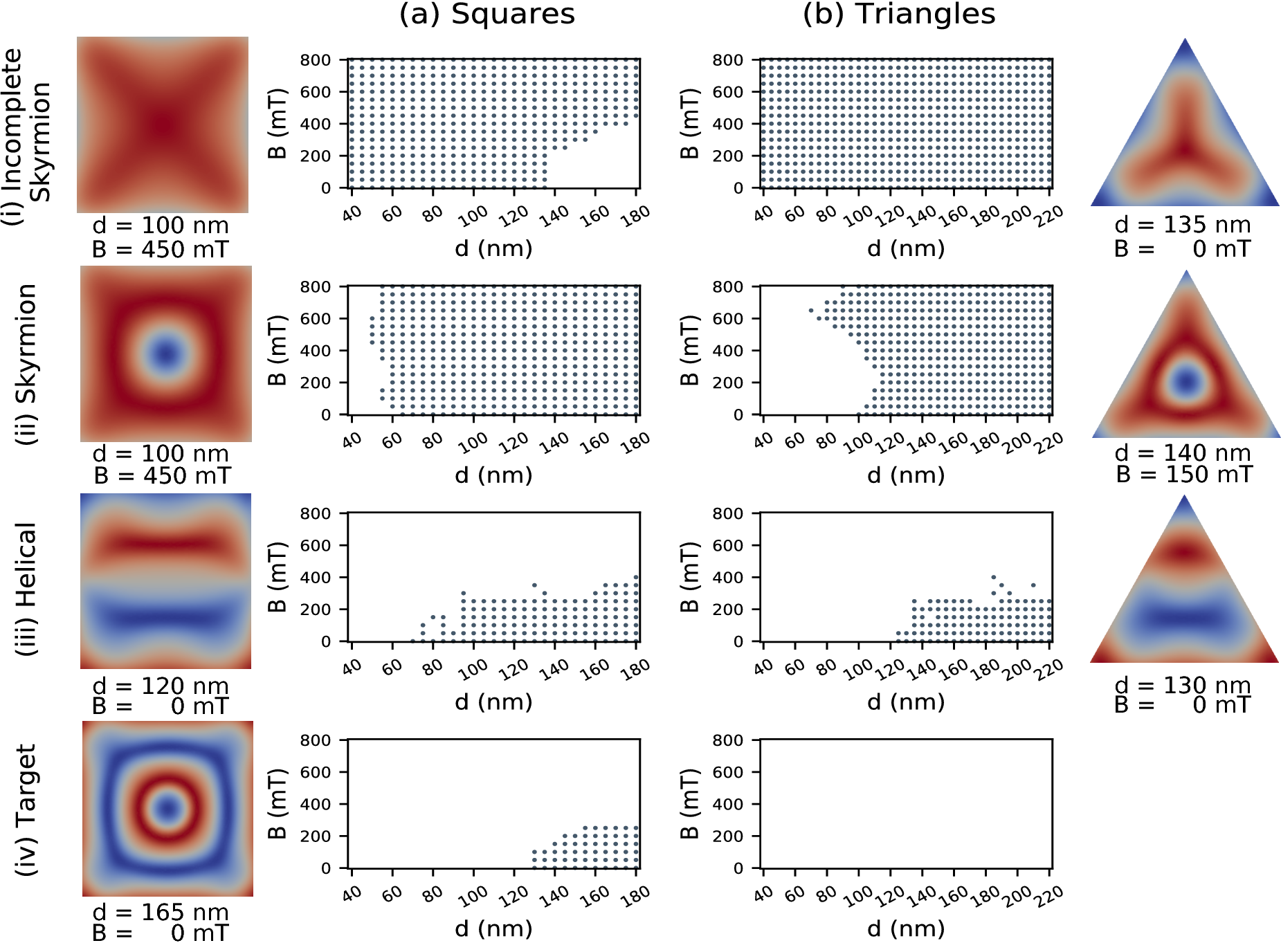}
\caption{Equilibrium regions for states found in different geometries. In the first and last columns we show the z-component of the magnetization for examples of states (i) to (iv) in the square and triangle systems. The $d$-$B$ graphs show dots when a metastable state of that type was found for that size and applied field. Incomplete skyrmion states (top row), are not stable for large square systems, with a field lower than around 400$\,$mT. However, this is noy seen in triangles. We note that in contrast to squares, we do not see target states as metastable in any region of phase space studied in triangles.}
\label{fig:metastable}
\end{figure*}
A wide variety of equilibrium states (formed of both the ground and metastable states) are obtained from the simulations in the systems, and in Fig.~\ref{fig:metastable}, we show the regions in $d-B$ phase space where each state can form as an equilibrium state. The equilibrium states can be broadly classified into several groups.
\begin{figure}[h!]
\centering
\includegraphics[width=0.9\linewidth]{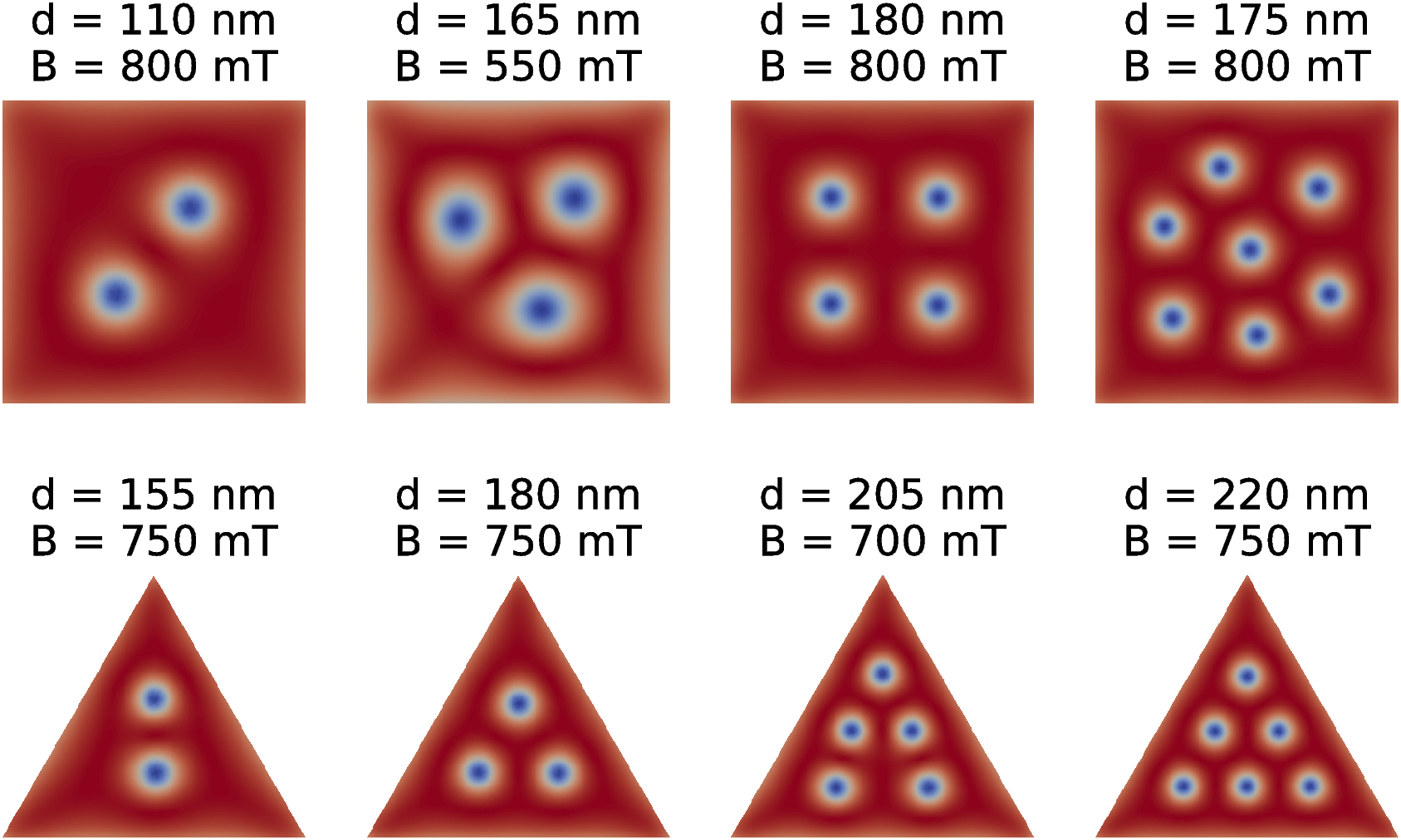}
\caption{Examples of the z-component of the magnetization for high energy states containing multiple skyrmions. For square systems, we saw states containing up to 10 skyrmions, and in triangles, a maximum of 6 skyrmions were observed.}
\label{fig:skyrmionzoo}
\end{figure}
\begin{enumerate}
\item \textit{Incomplete Skyrmions} - These states are named  \cite{Beg2015} as such due to the presence of a quasi-uniform magnetization across the system. Due to the DMI, at the boundaries we see twisting of the magnetization. (Fig.~\ref{fig:metastable} (i))
\item \textit{Isolated Skyrmions} These states, normally axially symmetric in disks, are distorted by the boundary of the confined geometry in both triangular and square systems. (Fig.~\ref{fig:metastable} (ii))
\item \textit{Helical States} A large variety of rotational spin textures form metastable states in the studied systems. (Fig.~\ref{fig:metastable} (iii))
\item \textit{Target States} Target states can be considered as an isolated skyrmion, with an additional radial half-helical rotation. (Fig.~\ref{fig:metastable} (iv))
\item \textit{Skyrmion Clusters} Multiple clusters of skyrmions form metastable states in the geometries when strong fields are applied to the system, resulting in a smaller skyrmion radius. We find these as high-energy metastable states for larger system sizes and for high applied fields. (Fig.~\ref{fig:skyrmionzoo})
\end{enumerate}
\begin{figure*}
\centering
\includegraphics[width=\figwit{}]{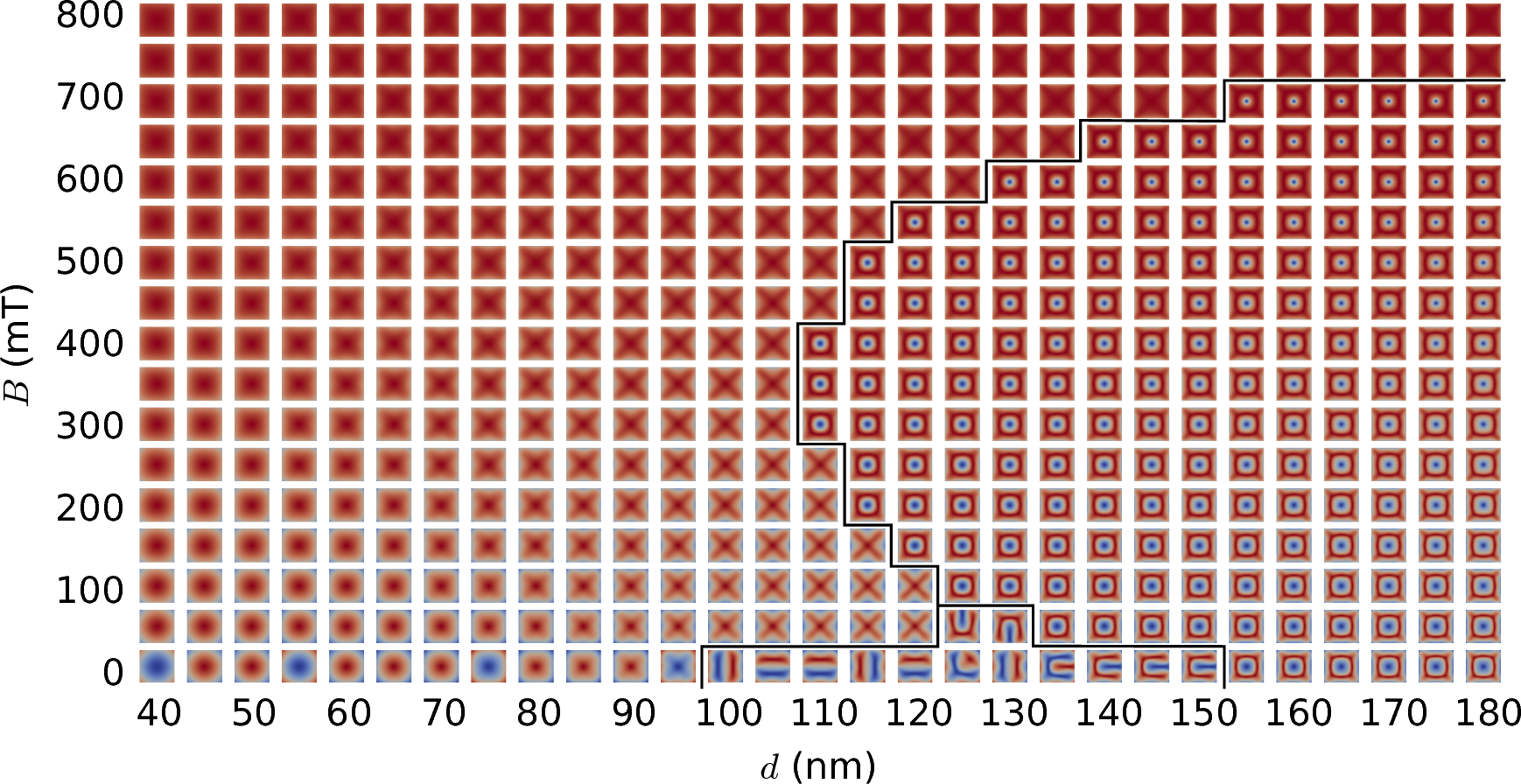}
\caption{In this $d-B$ ground state phase diagram for the square geometry, we show the z-component of the magnetization for the lowest energy state found for each sample size and applied field value. We see three regions of interest. (a)~The incomplete skyrmion state forms the bulk of the phase diagram. (b)~For a narrow region and for low applied field values, we see that helical states form the ground state. (c)~ Isolated skyrmions form the ground state for large sample sizes. As the field is increased, we can see the skyrmion shrinks such that more of the system aligns with the applied field, and for high field values the skyrmion no longer forms the ground state.}
\label{fig:squarephase}
\end{figure*}
\begin{figure*}
\centering
\includegraphics[width=\figwit{}]{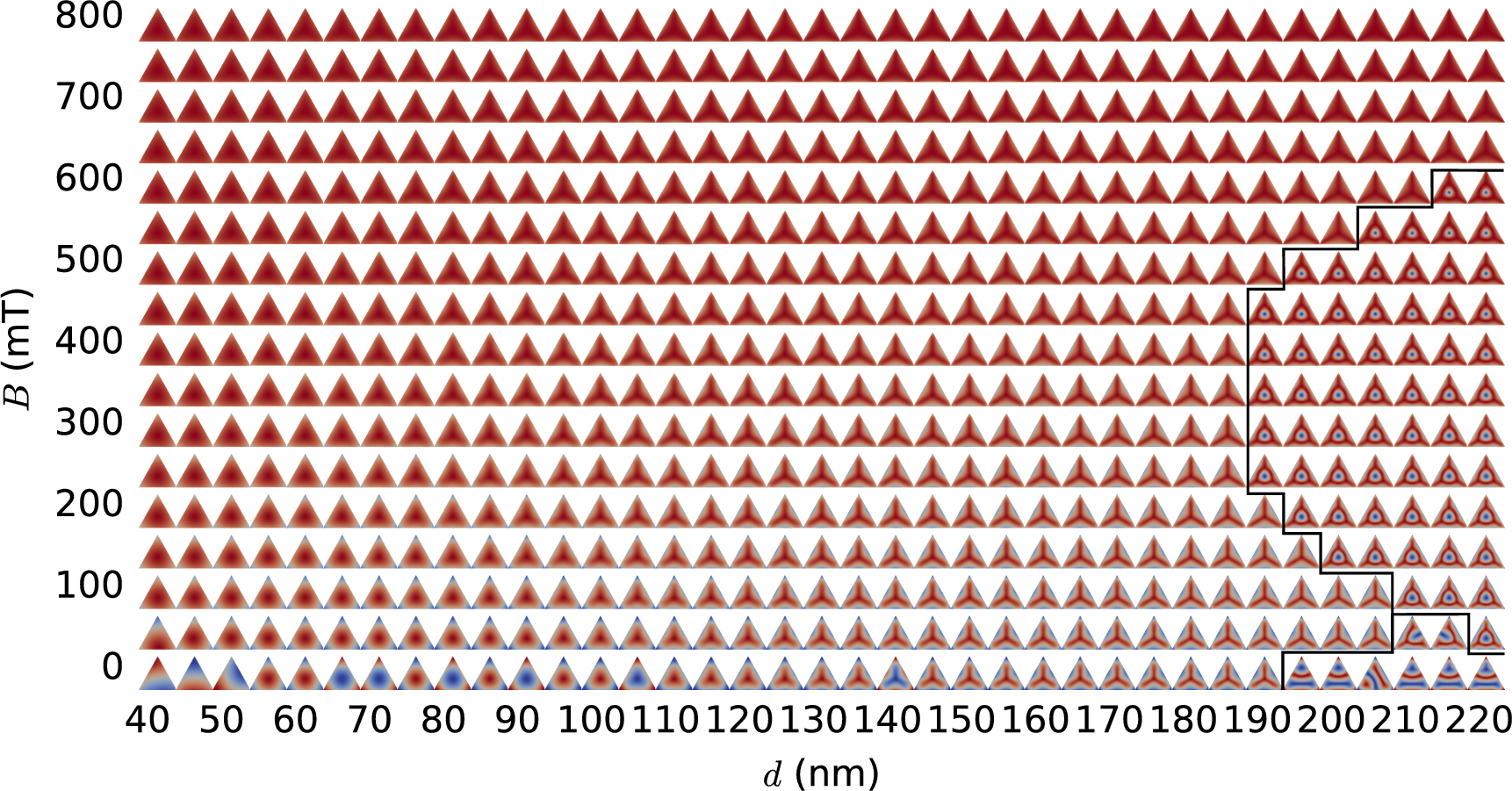}
\caption{Here, we show the $d-B$ ground state phase diagram for the triangle geometry, with the z-component of the magnetisation shown for each state obtained. As in Fig. \ref{fig:squarephase} we see three regions of interest. (a)~The incomplete skyrmion state forms the bulk of the phase diagram. (b)~For very large triangles, again at low applied field values, we see that helical states form the ground state. (c)~Skyrmions form the ground state for large sample sizes, but we do not observe these without an applied field.}
\label{fig:triphase}
\end{figure*}
\begin{figure*}
\centering
\includegraphics[width=\figwit{}]{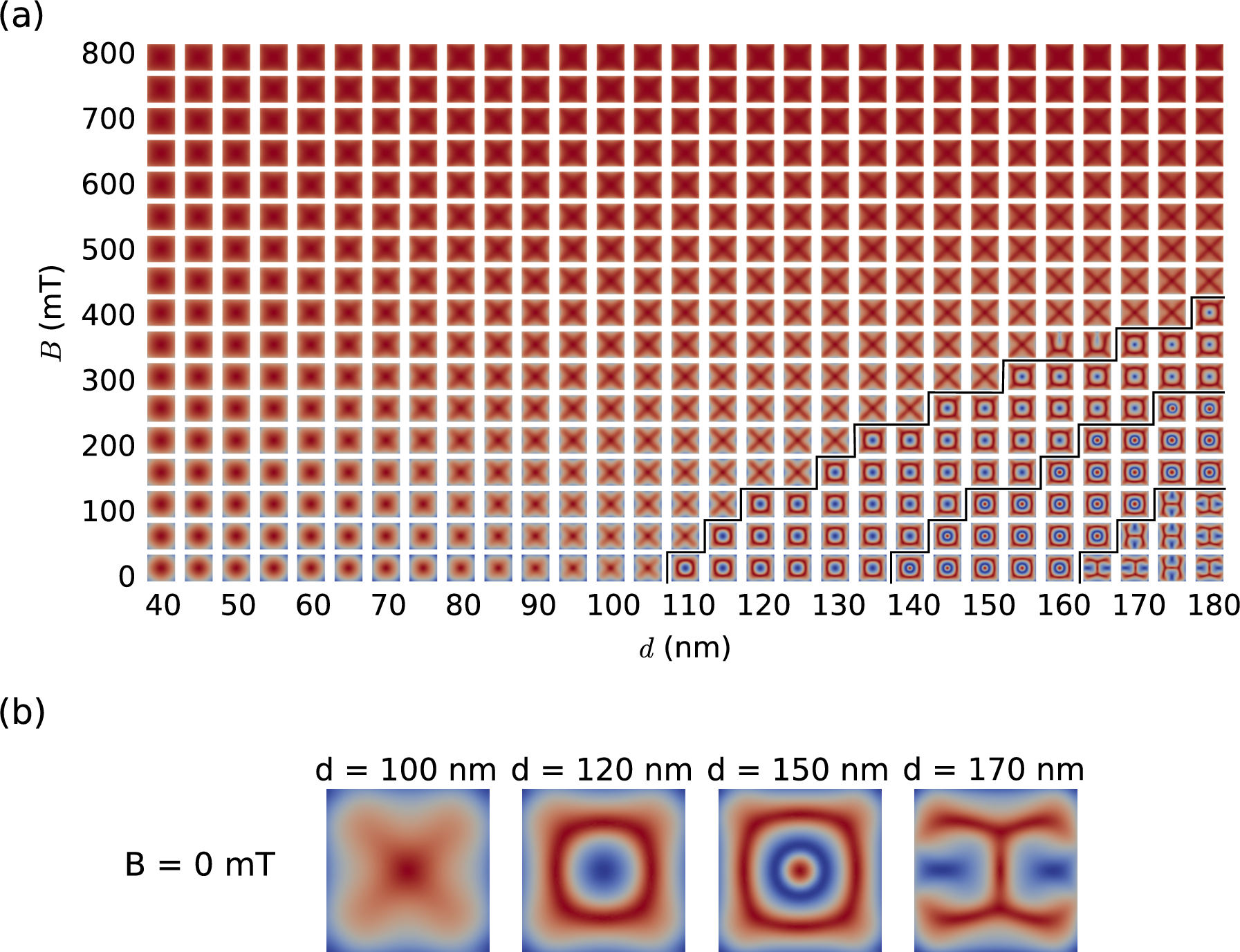}
\caption{Obtained states from relaxing square systems from uniform
magnetization. In (a), we see four regions of different type of states - from
left to right (i)~incomplete skyrmions (ii)~isolated skyrmions (iii)~overcomplete
skyrmions, and (iv)~helical type states. We predict that these configurations can be
achieved in an experimental study where first a high saturation field is applied
in the out-of-plane direction, and then the field is reduced to the value shown
on the $y$-axis. In (b), we show the final state obtained for each of these
configurations with no applied field.}
\label{fig:uniform-relaxed}
\end{figure*}
\subsection{Ground States}
\begin{figure*}
\centering
\includegraphics[width=\figwit{}]{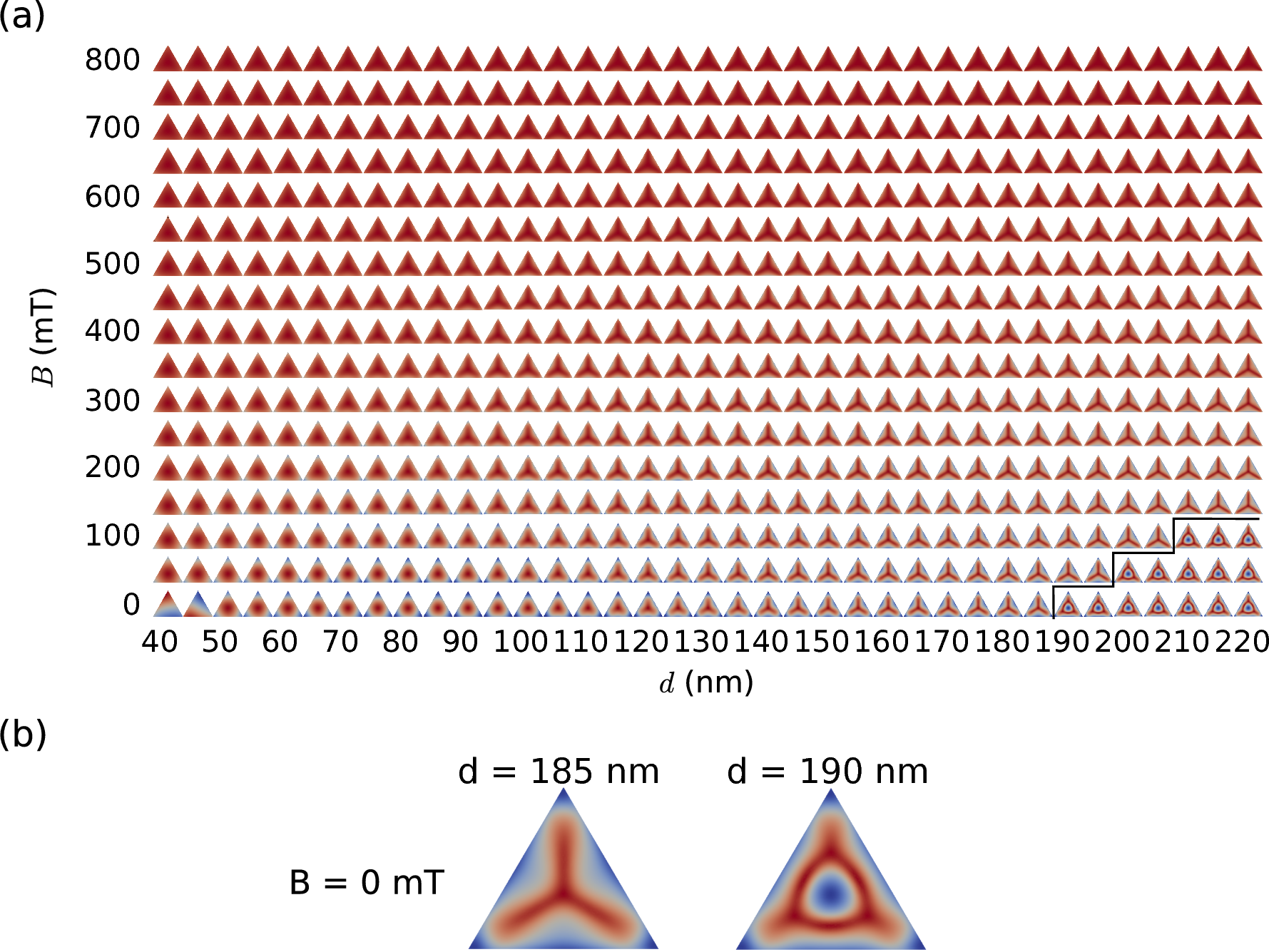}
\caption{Here we show the z-component of the magnetization for the states obtained by relaxing triangular systems from uniform
magnetization. In (a), we see two regions of different type of states - from
left to right (i)~incomplete skyrmions (ii)~isolated skyrmions. In (b), we show the final state obtained for each of these configurations with no applied field.}
\label{fig:uniform-relaxed-tri}
\end{figure*}
The ground state phase diagrams (Figs.~\ref{fig:squarephase} and \ref{fig:triphase})
show the lowest energy states identified for each geometry size for a given
applied field. For the square systems we see a large region where isolated
skyrmions form the lowest energy state for sample sizes as low as
$110\,\text{nm}$ with an applied field of 350$\,$mT. For larger sizes, the range
of applied fields where skyrmions form the ground state increases, and at
$155\,\text{nm}$, we compute that the skyrmion is the ground state with no
applied field. For all sample sizes studied, we see that applied fields of above
$700\,\text{mT}$ result in nearly uniform magnetization. With no applied field,
from $100\,\text{nm}$ to $150\,\text{nm}$ we see several types of helical states
form the ground state. These results are qualitatively similar to those seen in
disk systems, though in disks, skyrmions formed the ground state with no
applied field for smaller systems than in squares, with observation at disks of
diameter greater than $135\,\text{nm}$. \cite{Beg2015}

In Fig.~\ref{fig:triphase}, the ground state phase diagram for the triangular
systems is shown. In contrast to the square systems, we do not identify
skyrmions as the ground state when no applied field is applied for any sized sample we investigated, which shows a strong indication that the shape of the boundary of the system plays a crucial role in the energetics of magnetic skyrmions in confined geometries. Skyrmion states do form the ground state for systems of side length $d > 185\,\text{nm}$ when an applied field of between $50$ and $600\,\text{mT}$ is present. For systems of side length $d > 190\,\text{nm}$, we see a number of helical states form the ground state with no field. We note that between 40 and 50$\,$nm, we see quasi-helical type states, though the lengths in these systems are below the helical length of FeGe.

The incomplete skyrmion states identified in the triangular geometry vary
significantly depending on the size of the systems. Notably, tilting of the
magnetization at the boundary of the sample due to the DM interaction causes the
magnetization to point most strongly along the axis of the applied applied
field, with the strongest alignment along the axes of symmetry in both the
square and triangular states, which can be seen in the incomplete skyrmion
images shown in Fig.~\ref{fig:metastable}~(i).
\subsection{Proposed experimental study}
Of additional interest are states obtained from relaxing the systems from the
uniform state. Experimentally, these states could be realised by initially
applying a very strong applied field, to ensure that the magnetization of a
sample is saturated, and then rapidly reducing the applied magnetic field. The
states obtained from doing this in the square sample are shown in
Fig.~\ref{fig:uniform-relaxed}. We see four distinct bands of states, with
incomplete skyrmions forming the bulk of the phase diagram. Skyrmion states are
obtained in a narrow band, between $110$ and $135\,\text{nm}$, with no applied
field, and at larger sizes of system up to $180\,\text{nm}$ with an applied
field of $400\,\text{mT}$. For system sizes, from $140$ to $160\,\text{nm}$ with
no applied field, we see target states, and at $165\,\text{nm}$ and above with
no applied field, we identify helical states.

The corresponding uniform applied field results for triangular systems is shown
in Fig.~\ref{fig:uniform-relaxed-tri}. Here, we see similar results; in the bulk
of the phase diagram we see incomplete skyrmion states. For large systems of
between $190$ and $220\,\text{nm}$, when the field is reduced to a value between
$0$ and $150\,\text{mT}$, we see a small band of skyrmion states.
\section{Summary}
We show in this paper through micromagnetic simulations that in $10\,\text{nm}$ thick confined
geometries of FeGe, skyrmions can form the lowest energy state. When there is no
applied field, there exists a lower bound of side length $d$ between $150$ and $155\,\text{nm}$,
below which skyrmions do not form in square systems, and between
$100\,\text{nm}$ and $150\,\text{nm}$, a variety of helical type states form
the ground state. In triangular systems, we see the incomplete skyrmion state forms the ground state in most of
the phase space studied, and in large systems skyrmions form the ground
state between fields of $50$ and $600\,\text{mT}$.

We show over the same range of sizes and fields studied, a wide variety of states
are in equilibrium, and we show where these states can be obtained. We present the states obtained from relaxing uniformly
magnetised states from the saturated state, in both the square and triangular
systems in order to motivate experimental work on FeGe confined geometries, and
predict that skyrmion states should be experimentally accessible in both square
and triangular systems. We also predict that in large square systems, target
states should be accessible using the same procedure.
\section{Data Access Statement}
All data supporting this study are openly available from the Zenodo repository
at https://doi.org/10.5281/zenodo.1066791.
\section{Acknowledgements}
This work was financially supported by EPSRC Doctoral Training Centre
grant (EP/L015382/1), EPSRC Doctoral Training Centre Grant
EP/G03690X/1, OpenDreamKit Horizon 2020 European Research
Infrastructure project (676541), and the EPSRC Programme grant on
Skyrmionics (EP/N032128/1). D.C-O acknowledgees the financial support
from CONICYT Chilean scholarship programme Becas Chile (72130061). We
acknowledge the use of the University of Southampton IRIDIS High
Performance Computing Facility. T.K. acknowledges financial support from the Gordon and Betty Moore Foundation. W.W. acknowledges the financial support of the
National Natural Science Foundation of China (Grant No. 11604169).
\section{References}
%

\end{document}